\documentclass[prd,aps,showpacs,preprintnumbers,amssymb]{revtex4}
\usepackage{axodraw}
\usepackage{color}
\usepackage{epsf}

\def\e3p{$\eta \rightarrow 3 \pi$}

\begin{document}
\title{%
\hfill{\normalsize\vbox{%
\hbox{}
 }}\\
{Tree level relations in a composite electroweak theory}}

\author{Amir H. Fariborz
$^{\it \bf a}$~\footnote[1]{Email:
 fariboa@sunyit.edu}}
\author{Renata Jora
$^{\it \bf b}$~\footnote[2]{Email:
 rjora@theory.nipne.ro}}

\author{Joseph Schechter
 $^{\it \bf c}$~\footnote[4]{Email:
 schechte@phy.syr.edu}}

\affiliation{$^{\bf \it a}$ Department of Engineering, Science and Mathematics, State University of New York Institute of Technology, Utica, NY 13504-3050, USA}
\affiliation{$^{\bf \it b}$ National Institute of Physics and Nuclear Engineering PO Box MG-6, Bucharest-Magurele, Romania}

\affiliation{$^ {\bf \it c}$ Department of Physics,
 Syracuse University, Syracuse, NY 13244-1130, USA}

\date{\today}

\begin{abstract}
We work out simple tree level relations in a top condensate model with dynamical electroweak symmetry breaking. We find that in this picture the mass of the composite Higgs boson at tree level is given by $m_H^2=\frac{m_t^2}{2}$ where $m_t$ is the mass of the top quark.
\end{abstract}
\maketitle


Recently the ATLAS \cite{Atlas} and CMS \cite{CMS} experiments at CERN  have confirmed the existence  of a Higgs like particle with a mass $m_H=125.5$ GeV which has approximately \cite{Lin} all the properties of the standard model Higgs boson. The question remains if this Higgs boson is part of a larger structure like that of a supersymmetric model \cite{Ferretti} or of a model with dynamical symmetry breaking. The standard model by itself does not predict the mass of the Higgs  boson and one needs additional assumptions and extensions to determine this mass.

The Higgs boson can be composite in a large variety of models with dynamical symmetry breaking like the topcolor models \cite{Hill1},\cite{Hill2} or technicolor ones \cite{Sannino1}, \cite{Sannino2}.
In this work we will consider the simple case of a single composite Higgs doublet.  For simplicity we shall take the structure of this Higgs boson given by the top condensate models \cite{Bardeen}:

\begin{eqnarray}
\Phi=
\left[
\begin{array}{c}
t_R^{\dagger}b_L\\
t_R^{\dagger}t_L
\end{array}
\right].
\label{tsruct678}
\end{eqnarray}

In \cite{Jora} we explored the possibility  that by analogy with the low energy QCD in this kind of model also four quark states belonging to an additional Higgs doublet and a Higgs triplet might exist. We shall neglect this possibility here.
 Here we make the simple assumption that the bound state form through a Nambu Jona-Lasinio \cite{Nambu} mechanism and rely on the simplest top condensate model described in \cite{Bardeen}. For clarity we will resume to the electroweak sector of the standard model with only the top bottom quark doublet.

The model we wish to discuss has the initial Lagrangian:
\begin{eqnarray}
{\cal L}={\cal L}_{kin}+G(\bar{\Psi}_Lt_R\bar{t}_R\Psi_L)
\label{lagr5467}
\end{eqnarray}
where the kinetic term also includes the gauge interactions. It was shown in \cite{Bardeen}  that this Lagrangian generates a mass $m_t$ for the top quark, a composite scalar with the mass $m_s=2m_t$ and three massless pseudo Goldstone bosons. These states have the correct quantum numbers to lead to the dynamical breaking of the electroweak symmetry. The mass of the neutral scalar is obtained from the pole in the sum of the scalar channel fermion bubble and it is a definite prediction of this type of models. This prediction might be improved by considering corrections coming from the gauge sector and from higher order fermion operators but even in this case the predicted mass of the composite scalar is  at least two times the mass of the Higgs boson found at the LHC.

One might take as an example another Lagrangian that of a linear sigma model describing the light mesons in low energy QCD \cite{Schechter} which in its simplest form is:
\begin{eqnarray}
{\cal L}=-\frac{1}{2}{\rm Tr}[\partial^{\mu}M\partial_{\mu}M^{\dagger}]-c_2{\rm Tr}[MM^{\dagger}]+2 {\rm Tr} [AS]
\label{linerasigmsmodel}
\end{eqnarray}
where the $M$ field represents an octet of scalar and pseudoscalar mesons and the last term breaks explicitly the chiral symmetry and it is proportional to the mass of the light quarks (Note that the actual Lagrangian in \cite{Schechter} is more complex but we truncated it for the purpose of illustration). It turns out that the sigma in the simple model described in Eq. (\ref{linerasigmsmodel}) has a mass $m_s^2=\frac{2A}{\alpha}$ (for simplicity we took the symmetry  breaking term degenerate) so the fact that the composite state has  a mass dependent directly on the mass of the elementary quarks is quite generic.


The Lagrangian in Eq. (\ref{lagr5467}) leads, at an energy lower that  of formation of composite states, to the effective Lagrangian:
\begin{eqnarray}
{\cal L}={\cal L}_{kinetic}+g_t(\bar{\Psi}_Lt_RH+h.c.)+Z_H(D_{\mu}H)^2-m_H^2H^{\dagger}H-\frac{\lambda}{2}(H^{\dagger}H)^2.
\label{effl879}
\end{eqnarray}

Using as a start-up this Lagrangian and the standard model renormalization group equations one can predict \cite{Bardeen} the mass of the top quark and of the Higgs boson in a more precise way.

Here however we shall adopt a different point of view and consider instead the Lagrangian at the scale where both elementary and composite states exist and the latter takes the form:
\begin{eqnarray}
{\cal L}'={\cal L}_{kinetic}+g_t(\bar{\Psi}_Lt_RH+h.c.)+(D_{\mu}H)^2-m_H^2H^{\dagger}H-\frac{\lambda}{2}(H^{\dagger}H)^2
\label{newlagr567}
\end{eqnarray}
We assume that the above structure makes sense also in terms of the quark elementary states and that both the equation of motion of the elementary quarks and of the Higgs boson are respected. We express the composite Higgs boson as:
\begin{eqnarray}
H_0=\frac{1}{\Lambda^2}{\rm Re}[t_R^{\dagger}t_L]=\frac{1}{2\Lambda^2}(\bar{t}t)
\label{repr890}
\end{eqnarray}

Let us see what is the naive mass of the Higgs bosons in terms of its constituents. One knows that nonrelativistically in the center mass frame the mass of the bound state is simply $2m_t$ which confirms the previous findings. In the relativistic case the equation of motion is:
\begin{eqnarray}
\partial^{\mu}\partial_{\mu}H_0+m_H^2H_0=0,
\label{rez89}
\end{eqnarray}
which translated for the composite states (we took generically for these $\bar\Psi\Psi$) has the form:
 \begin{eqnarray}
&&(\partial^{\mu}\partial_{\mu}\bar{\Psi})\Psi+\bar{\Psi}(\partial^{\mu}\partial_{\mu}\Psi)+2\partial^{\mu}\bar{\Psi}\partial_{\mu}\Psi+m_H^2\bar{\Psi}\Psi=0
\nonumber\\
&&-m_t^2\bar{\Psi}\Psi-m_t^2\bar{\Psi}\Psi-2\partial^{\mu}\bar{\Psi}\gamma^{\mu}\gamma^{\rho}\partial_{\rho}\Psi-m_H^2\bar{\Psi}\Psi=
(-4m_t^2+m_H^2)\bar\Psi\Psi=0,
\label{rez67589}
\end{eqnarray}
where $m_t$ is the mass of the top quark.
Here we applied the equation of motion for the constituents quarks and also used:
\begin{eqnarray}
\frac{1}{2}\{\gamma^{\mu},\gamma^{\rho}\}=\gamma^{\mu}\gamma^{\rho}-\frac{1}{2}[\gamma^{\mu},\gamma^{\rho}],
\label{rez6789}
\end{eqnarray}
and the fact that in the Fourier space:
\begin{eqnarray}
&&\partial^{\mu}\bar{\Psi}[\gamma^{\mu},\gamma^{\rho}]\partial_{\rho}\Psi=
\int\frac{d^4k}{(2\pi)^4}k^{\mu}(r+k)_{\rho}\bar{\Psi}(-k)[\gamma^{\mu},\gamma^{\rho}]\Psi(k+r)=\int\frac{d^4k}{(2\pi)^4}k_{\rho}(r+k)^{\mu}\bar{\Psi}(k+r)[\gamma^{\mu},\gamma^{\rho}]\Psi(-k)
\nonumber\\
&&=\int\frac{d^4k}{(2\pi)^4}k^{\mu}r_{\rho}\bar{\Psi}(-k)[\gamma^{\mu},\gamma^{\rho}]\Psi(k+r)=\int\frac{d^4k}{(2\pi)^4}k_{\rho}r^{\mu}\bar{\Psi}(k+r)[\gamma^{\mu},\gamma^{\rho}]\Psi(-k),
\nonumber\\
&&\int\frac{d^4k}{(2\pi)^4}k^{\mu}r_{\rho}\bar{\Psi}(-k)[\gamma^{\mu},\gamma^{\rho}]\Psi(k+r)=-\int\frac{d^4k}{(2\pi)^4}k^{\mu}r_{\rho}\bar{\Psi}(-k)[\gamma^{\mu},\gamma^{\rho}]\Psi(k+r)=0.
\label{rez645789}
\end{eqnarray}
In the r.h.s of the last line of the equation made again the change of variable $-k'=k+r$.

Thus the result $m_H^2=4m_t^2$ is a result quite valid also from this point of view of the constituent quark equation of motion.


Let us consider a simplified version of the Lagrangian in Eq. (\ref{newlagr567}) which does not contain the Higgs kinetic and quartic term. The Higgs can be integrated out and this leads to:
\begin{eqnarray}
H_0=\frac{g_t}{m_H^2}{\rm Re}[t_R^{\dagger}t_L]=\frac{g_t}{2m_H^2}\bar{t}t
\label{higg678}
\end{eqnarray}

Comparing Eq. (\ref{repr890}) with Eq. (\ref{higg678}) we determine that $\Lambda^2=\frac{m_H^2}{g_t}$. In what follows we will use the expression in Eq. (\ref{higg678}) for the composite Higgs.

We return to the full Lagrangian in Eq. (\ref{newlagr567}) and regard this in term of the elementary quarks. One might wonder if we preserve the equation of motion in first order for the fermion fields what does the addition of the Higgs kinetic term and of the four quark terms adds to the picture. To do that we first work out:
\begin{eqnarray}
g_t(t_L^{\dagger}t_RH+h.c.)-m_H^2H^{\dagger}H=\frac{g_t^2}{m_H^2}(t_L^{\dagger}t_Rt_R^{\dagger}t_L+h.c)-\frac{g_t^2}{m_H^2}(t_L^{\dagger}t_Rt_R^{\dagger}t_L)=
\frac{g_t^2}{4m_H^2}\bar{t}t\bar{t}t
\label{term657}
\end{eqnarray}

Now let us apply the equation of motion to the full Lagrangian:
\begin{eqnarray}
{\cal L}=i\bar{t}\gamma^{\mu}\partial_{\mu}t+\frac{g_t^2}{4m_H^4}\partial^{\mu}(\bar{t}t)\partial_{\mu}(\bar{t}t)+\frac{g_t^2}{4m_H^2}\bar{t}t\bar{t}t-\frac{\lambda}{2}\frac{g_t^4}{16m_H^8}(\bar{t}t)^4,
\label{ferm4567}
\end{eqnarray}

to get:
\begin{eqnarray}
-i\gamma^{\mu}\partial_{\mu}t+\frac{2g_t^2}{4m_H^4}t\partial_{\mu}\partial^{\mu}(\bar{t}t)-\frac{2g_t^2}{4m_H^2}t(\bar{t}t)+\frac{2g_t^2}{4m_H^2}t(\bar{t}t)=0
\label{masferm567}
\end{eqnarray}

We denote:
\begin{eqnarray}
\frac{g_t}{m_H^2}\langle \bar{t}t \rangle=2v
\label{vac4536}
\end{eqnarray}
and reinforce $m_t=g_tv$, $v^2=\frac{m_H^2}{\lambda}$ and the equation of motion for $\bar{t}t$ (see Eq. (\ref{rez67589})) to obtain for the effective mass of the top quark:
\begin{eqnarray}
m_t'=\frac{2g_t}{4m_H^4}2vm_H^2(-4m_t^{2\prime})=m_t\frac{m_t^{2\prime}}{m_t}
\label{mas5789}
\end{eqnarray}

The obvious solution is $m_t'=m_t$ so the mass of the quark is reinforced if we use $m_H^2=4m_t^2$.
However  the same Lagrangian that has a mass $m_H^2=4m_t^2$ leads after taking into consideration the vacuum expectation value of the field to a real mass $m_H^{2\prime}=-m_H^2+3m_H^2=2m_H^2$. This is obtained by adding the contribution of the quartic term as in the classic case of spontaneous symmetry breaking. We claim that this is inconsistent as  the mass obtained from both the equation of motion of the elementary and composite states should be the same. In order to fix that we need to take:
\begin{eqnarray}
&&2m_H^2=4m_t^2
\nonumber\\
&&m_H^2=2m_t^2
\label{tre6788}
\end{eqnarray}

We introduce this new value for $m_H^2$ in Eq. (\ref{masferm567}) and reinforce the equation of motion for the scalar bound state to get:
\begin{eqnarray}
m_t'=\frac{2g_t}{4m_H^4}2vm_H^2(-4m_t^{2})=-2g_tv
\label{term678}
\end{eqnarray}

We claim that this is the real tree level mass of the top quark in this model and in  term of this the Higgs boson mass is:
\begin{eqnarray}
m_H^2=2m_t^2=\frac{m_t^{2\prime}}{2}.
\label{true54678}
\end{eqnarray}

This argument works both ways. We can consider,
\begin{eqnarray}
2m_H^2=4m_t^{\prime 2},
\label{rez678}
\end{eqnarray}
where $m_t$ is the real mass of the top quark to get from Eq. (\ref{mas5789}):
\begin{eqnarray}
&&m_{t}^{\prime}=2m_t\frac{m_t^{\prime 2}}{m_t^2}
\nonumber\\
&&m_t^{\prime}=\frac{m_t}{2},
\label{som789}
\end{eqnarray}
to get again for the mass of the Higgs boson in terms of the real mass of the top quark $m_t$:
\begin{eqnarray}
m_H^2=2m_t^{2\prime}=\frac{m_t^2}{2}.
\label{real767}
\end{eqnarray}

With a mass of the top quark $m_{top}=173.07$ GeV one obtains for the tree level mass of the composite Higgs boson $M_H=122.4$ GeV which means that the quantum correction should contribute only with 3 percents to the mass of the Higgs boson.

In conclusion we argue that in a model with a top composite Higgs particle the structure of the Lagrangian and the equation of motion suggest that in first instance and at tree level the mass of the top quarks in the bound state is smaller than that of the free top quark. Of course for a precise determination one needs to consider the effect of the gauge bosons and of the loop corrections. This finding is different than the low energy QCD picture where it is observed that the mass of the quarks in the bound state, the " constituent" quark mass is higher than that of the free quark. However one should note between the two cases a striking difference. In the case of the low energy QCD in the vast majority of models the existence of the bound states mesons excludes the presence of the elementary quarks in the same Lagrangian. This is because the composite and elementary states belong to different energy scales. In contrast for the more intricate case of top composite Higgs model  the simple structure of the standard model Lagrangain suggests that the bound state and the free elementary quark coexist at about the same   energy scale. This indicates that this kind of models may have features that are different from those of low energy QCD.

In the end the slice of the standard model that contains the top quark and the Higgs bosons in the composite picture should have the Lagrangian:
 \begin{eqnarray}
{\cal L}={\cal L}_{kin}+g_t(\Psi_Lt_RH+h.c.)+|D_{\mu}H|^2+m^2H^{\dagger}H-\frac{\lambda_0}{2}(H^{\dagger}H)^2.
\label{st5467}
\end{eqnarray}
with the vacuum expectation value given by $v^2=\frac{m^2}{\lambda_0}$ and with a tree level mass of the Higgs boson $m_H^2=2m^2=\frac{m_t^2}{2}$ where $m_t=g_tv$.

\section*{Acknowledgments} \vskip -.5cm
The work of R. J. was supported by a grant of the Ministry of National Education, CNCS-UEFISCDI, project number PN-II-ID-PCE-2012-4-0078.


\begin{thebibliography}{15}
\bibitem{Atlas}  The ATLAS Collaboration, Phys. Lett. B {\bf 716}, 1 (2012).
\bibitem{CMS} The CMS Collaboration, Phys. Lett. B {\bf 716}, 30 (2012).
\bibitem{Lin} X.-G.He, S.-F.Li and H.-H.Lin, Mod. Phys. Lett. A {\bf 28}, 1350085 (2013), arXiv:1302.6302.
\bibitem{Ferretti} G. Ferretti and D. Karateev, Mod. Phys. Lett. A {\bf 28}, 1350025 (2013), arXiv:1206.0761.
\bibitem{Hill1} C. Hill, Phys. Lett. B {\bf 266}, 419-424 (1991)
\bibitem{Hill2} C. Hill, Phys. Lett. B {\bf 345}, 483-489 (1995), arXiv:hep-ph/9411426.
\bibitem{Sannino1} C. Hill and E. H. Simmons, Phys. Rep. {\bf 381}, 235-402 (2002), arXiv:hep-ph/0203079.
\bibitem{Sannino2} O. Antipin, F. Sannino and K. Tuominen, Mod. Phys. Lett. A {\bf 28}, 1350140 (2013), arXiv:1306.6514.
\bibitem{Bardeen} W. A. Bardeen, C. T. Hill and M. Lindner, Phys. Rev. D {\bf 41}, 1647 (1990).
\bibitem{Jora} R. Jora, S. Nasri and J. Schechter, Phys. Rev. D {\bf 87}, 115027 (2013), arXiv:1304.7139.
\bibitem{Nambu} Y. Nambu and G. Jona-Lasinio, Phys. Rev. {\bf 122}, 345-358 (1961); Phys. Rev. {\bf 124}, 246-254 (1961).
\bibitem{Schechter} A. H. Fariborz, R. Jora and J. Schechter, Phys. Rev. D {\bf 79}, 074014 (2009), arXiv:0902.2825.
\end{thebibliography}
\end{document}